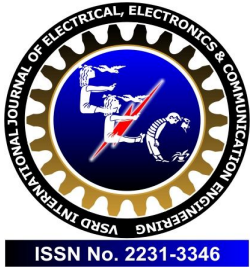

**RESEARCH COMMUNICATION**

# Smart Antenna for Cellular Mobile Communication

[1]RK Jain*, [2]Sumit Katiyar and [3]NK Agrawal

## ABSTRACT

The adoption of smart / adaptive antenna techniques in future wireless systems is expected to have a significant impact on the efficient use of the spectrum, the minimization of the cost of establishing new wireless networks, the optimization of service quality and realization of transparent operation across multi technology wireless networks [1]. This paper presents brief account on smart antenna (SA) system. SAs can place nulls in the direction of interferers via adaptive updating of weights linked to each antenna element. SAs thus cancel out most of the co-channel interference resulting in better quality of reception and lower dropped calls. SAs can also track the user within a cell via direction of arrival algorithms [2]. This paper explains the architecture, evolution and how the smart / adaptive antenna differs from the basic format of antenna. The paper further explains about the radiation pattern of the antenna and why it is highly preferred in its relative field. The capabilities of smart / adaptive antenna are easily employable to Cognitive Radio and OFDMA system.

## 1. INTRODUCTION

In view of explosive growth in the number of digital cellular subscribers, service providers are becoming increasingly concerned with the limited capacities of their existing networks. This concern has led to the deployment of smart antenna systems throughout major metropolitan cellular markets. These smart antenna systems have typically employed multibeam technologies, which have been shown, through extensive analysis, simulation, and experimentation, to provide substantial performance improvements in FDMA, TDMA and CDMA networks [3-7]. Multibeam architectures for FDMA and TDMA systems provide the straight-forward ability of the smart antenna to be implemented as a non-invasive add-on or appliqué to an existing cell site, without major modifications or special interfaces [22].

This paper mainly concentrates on use of smart antennas in mobile communications that enhances the

---
[1,2]Research Scholar, Department of Electrical & Electronics Engineering, Singhania University, Jhunjhunu, Rajasthan, INDIA. [3]Professor, Department of Electronics & Communication Engineering, Inderprastha Engineering College, Ghaziabad, Uttar Pradesh, INDIA. *Correspondence : rkjain_iti@rediffmail.com

capabilities of the mobile and cellular system such as faster bit rate, multi use interference, space division multiplexing (SDMA), adaptive SDMA [21], increase in range, multipath mitigation, reduction of errors due to multipath fading, best suitability of multi-carrier modulations such as OFDMA. The best application of SAs is its suitability for demand based frequency allocation in hierarchical system approach (flexible antenna pattern are achieved electronically and no physical movement of receiving antennas is necessary). The advantage of SAs application in cellular systems are decreased inter symbol interference, decreased co-channel interference & adjacent channel interference, improved bit error rate (due to decreased amount of multipath and ISI), increase in receiver sensitivity, reduction in power consumption & RF pollution. Smart antennas are most appropriate for use of cognitive radio (software radio technology provides flexibility) and the greatest advantage of smart antenna is a very high security.

The main impediments to high-performance wireless communications are interference from other users (co-channel interference), the inter-symbol interference (ISI) and signal fading caused by multipath. Co-channel interference limits the system capacity, defined as the number of users which can be serviced by the system. However, since the desired signal and co-channel interference typically arrive at the receiver from different directions, smart antennas can exploit these differences to reduce co-channel interference, thereby increasing system capacity. The reflected multipath components of the transmitted signal also arrive at the receiver from different directions, and spatial processing can use these differences to attenuate the multipath, thereby reducing ISI and fading. Since data rate and BER are degraded by these multipath effects, reduction in multipath through spatial processing can lead to higher data rates and better BER performance.

In a cellular system, omni-directional antennas have traditionally been used at base stations to enhance the coverage area of the base stations but it also leads a gross wastage of power that in-fact is the main cause of co-channel interference at neighboring base stations. The sectoring concept with diversity system exploits space diversity and results in improve reception by counteracting with negative effects of multipath fading. Adaptive / smart antenna technology represents the most advanced smart antenna approach to date. Using a variety of new signal-processing algorithms, the adaptive system takes advantage of its ability to effectively locate and track various types of signals to dynamically minimize interference and maximize intended signal reception. Both adaptive / smart systems attempt to increase gain according to the location of the user; however; only the adaptive system provides optimal gain while simultaneously identifying, tracking, and minimizing interfering signals.

## 2. SMART ANTENNA

### 2.1. Smart

The concept of using multiple antennas and innovative signal processing to serve cells more intelligently has existed for many years. In fact, varying degrees of relatively costly smart antenna systems have already been applied in defense systems. Until recent years, cost barriers have prevented their use in commercial systems. The advent of powerful low-cost digital signal processors (DSPs), general-purpose processors (and ASICs), as well as innovative software-based signal-processing techniques (algorithms) have made intelligent antennas practical for cellular communications systems.

Today, when spectrally efficient solutions are increasingly a business imperative, these systems are providing greater coverage area for each cell site, higher rejection of interference, and substantial capacity improvements.

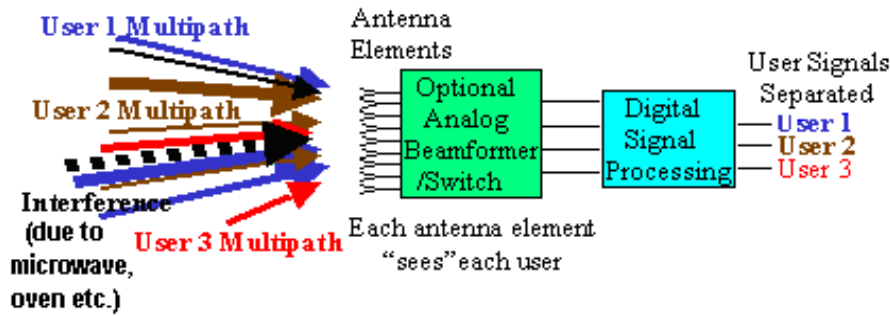

**Fig. 1 : Smart Antenna System**

## 2.2. What Is a Smart Antenna System?

In truth, antennas are not smart antenna systems are smart. Generally co-located with a base station, a smart antenna system combines an antenna array with a digital signal-processing capability to transmit and receive in an adaptive, spatially sensitive manner. Such a configuration dramatically enhances the capacity of a wireless link through a combination of diversity gain, array gain and interference suppression. Increased capacity translates to higher data rates for a given number of users or more users for a given data rate per user. In other words, such a system can automatically change the directionality of its radiation patterns in response to its signal environment. This can dramatically increase the performance characteristics (such as capacity) of a wireless system.

Multipath of propagation are created by reflections and scattering. Also, interference signals such as that produced by the microwave oven in the picture fig (1) are superimposed on the desired signals. Measurements suggest that each path is really a bundle or cluster of paths, resulting from surface roughness or irregularities. The random gain of the bundle is called multipath fading [7], [8], [9], [10].

## 2.3. How Many Types of Smart Antenna Systems Are There?

Terms commonly heard today that embrace various aspects of a smart antenna system technology include intelligent antennas, phased array, SDMA, spatial processing, digital beam forming, adaptive antenna systems, and others. Smart antenna systems are customarily categorized, however, as either switched beam or adaptive array systems.

The following are distinctions between the two major categories of smart antennas regarding the choices in transmit strategy:

• Switched Beam—a finite number of fixed, predefined patterns or combining strategies (sectors)

• Adaptive Array—an infinite number of patterns (scenario-based) that are adjusted in real time

### 2.3.1.   What Are Switched Beam Antennas?

Switched beam antenna systems form multiple fixed beams with heightened sensitivity in particular directions. These antenna systems detect signal strength, choose from one of several predetermined, fixed beams, and switch from one beam to another as the mobile moves throughout the sector.

Instead of shaping the directional antenna pattern with the metallic properties and physical design of a single element (like a sectorized antenna), switched beam systems combine the outputs of multiple antennas in such a way as to form finely sectorized (directional) beams with more spatial selectivity than can be achieved with conventional, single-element approaches fig (2).

### 2.3.2.   What Are Adaptive Array Antennas?

Adaptive antenna technology represents the most advanced smart antenna approach to date. Using a variety of new signal-processing algorithms, the adaptive system takes advantage of its ability to effectively locate and track various types of signals to dynamically minimize interference and maximize intended signal reception.

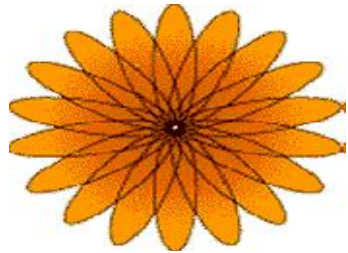

**Fig.  2 :  Switched Beam System Coverage Patterns (Sectors)**

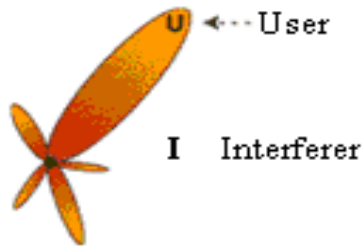

**Fig. 3 :  Adaptive Array Antenna**

Both systems attempt to increase gain according to the location of the user; however, only the adaptive system provides optimal gain while simultaneously identifying, tracking, and minimizing interfering signals.

Adaptive Array Coverage: A representative depiction of a main lobe extending toward a tser with a null directed toward a co-channel interferer as shown in fig (3).

### 2.3.3.   What Do They Look Like?

Omni-directional antennas are obviously distinguished from their intelligent counterparts by the number of antennas (or antenna elements) employed. Switched beam and adaptive array systems, however, share many hardware characteristics and are distinguished primarily by their adaptive intelligence.

To process information that is directionally sensitive requires an array of antenna elements (typically 4 to 12), the inputs from which are combined to control signal transmission adaptively. Antenna elements can be arranged in linear, circular, or planar configurations and are most often installed at the base station, although they may also be used in mobile phones or laptop computers [19].

### 2.3.4. What Makes Them So Smart?

A simple antenna works for a simple RF environment. Smart antenna solutions are required as the number of users, interference, and propagation complexity grow. Their smarts reside in their digital signal-processing facilities. Like most modern advances in electronics today, the digital format for manipulating the RF data offers numerous advantages in terms of accuracy and flexibility of operation. Speech starts and ends as analog information. Along the way, however, smart antenna systems capture, convert, and modulate analog signals for transmission as digital signals and reconvert them to analog information on the other end. In adaptive antenna systems, this fundamental signal-processing capability is augmented by advanced techniques (algorithms) that are applied to control operation in the presence of complicated combinations of operating conditions. The benefit of maintaining a more focused and efficient use of the system's power and spectrum allocation can be significant [19].

## 3. THE GOALS OF A SMART ANTENNA SYSTEM

The dual purpose of a smart antenna system is to augment the signal quality of the radio-based system through more focused transmission of radio signals while enhancing capacity through increased frequency reuse.

**Table 1 :  Features and Benefits of Smart Antenna Systems [19].**

| Feature | Benefit |
|---|---|
| Signal Gain—Inputs from multiple antennas are combined to optimize available power required to establish given level of coverage. | Better Range / Coverage—Focusing the energy sent out into the cell increases base station range and coverage. Lower power requirements also enable a greater battery life and smaller/lighter handset size. |
| Interference Rejection—Antenna pattern can be generated toward cochannel interference sources, improving the signal-to-interference ratio of the received signals. | Increased Capacity—Precise control of signal nulls quality and mitigation of interference combine to frequency reuse reduce distance (or cluster size), improving capacity. Certain adaptive technologies (such as space division multiple access) support the reuse of frequencies within the same cell. |
| Spatial Diversity—Composite information from the array is used to minimize fading and other undesirable effects of multipath propagation. | Multipath Rejection—can reduce the effective delay spread of the channel, allowing higher bit rates to be supported without the use of an equalizer, improved bit error rate (due to decreased amount of multipath and ISI). |
| SDMA- SDMA continually adapts to the radio environment through intelligent / smart antenna. | Providing each user with uplink and downlink signals of the highest possible quality and can adapt the frequency allocation to where the most users are located. |
| Power Efficiency—combines the inputs to multiple elements to optimize available processing gain in the downlink (toward the user) | Reduced Expense—Lower amplifier costs, power consumption, and higher reliability will result. Lower power consumption reduces not only interferences but also reduces RF pollution (ease health hazard). It will also result in reduction of scares energy resource (diesel consumption) and save foreign currency. |

## 4. The Architecture of Smart Antenna Systems

### 4.1. How Do Smart Antenna Systems Work?

Traditional switched beam and adaptive array systems enable a base station to customize the beams they generate for each remote user effectively by means of internal feedback control. Generally speaking, each approach forms a main lobe toward individual users and attempts to reject interference or noise from outside of the main lobe.

### 4.2. Switched Beam Systems

In terms of radiation patterns, switched beam is an extension of the current microcellular or cellular sectorization method of splitting a typical cell. The switched beam approach further subdivides macro-sectors into several microsectors as a means of improving range and capacity. Each micro-sector contains a predetermined fixed beam pattern with the greatest sensitivity located in the center of the beam and less sensitivity elsewhere. The design of such systems involves high-gain, narrow azimuthal beam width antenna elements (fig 4).

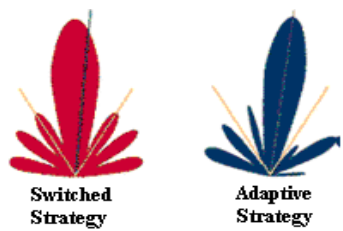

**Fig. 4 :  Beam forming Lobes and Nulls that Switched Beam (Red) and Adaptive Array (Blue) Systems Might Choose for Identical User Signals (Green Line) and Co-channel Interferers (Yellow Lines).**

Smart antenna systems communicate directionally by forming specific antenna beam patterns. When a smart antenna directs its main lobe with enhanced gain in the direction of the user, it naturally forms side lobes and nulls or areas of medium and minimal gain respectively in directions away from the main lobe.

Different switched beam and adaptive smart antenna systems control the lobes and the nulls with varying degrees of accuracy and flexibility.

### 4.3. Adaptive Antenna Approach

The adaptive antenna systems approach communication between a user and base station in a different way, in effect adding a dimension of space. By adjusting to an RF environment as it changes (or the spatial origin of signals), adaptive antenna technology can dynamically alter the signal patterns to near infinity to optimize the performance of the wireless system.

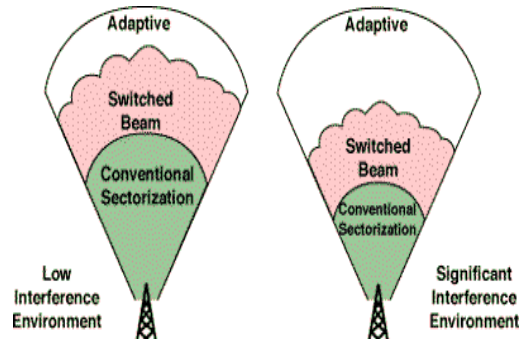

**Fig. 5 : Coverage Patterns for Switched Beam and Adaptive Array Antennas**

Adaptive arrays utilize sophisticated signal-processing algorithms to continuously distinguish between desired signals, multipath, and interfering signals as well as calculate their directions of arrival. This approach continuously updates it's transmit strategy based on changes in both the desired and interfering signal locations. The ability to track users smoothly with main lobes and interferers with nulls ensures that the link budget is constantly maximized because there are neither microsectors nor predefined patterns (fig 5).

Both types of smart antenna systems provide significant gains over conventional sectored systems. The low level of interference on the left represents a new wireless system with lower penetration levels. The significant level of interference on the right represents either a wireless system with more users or one using more aggressive frequency reuse patterns. In this scenario, the interference rejection capability of the adaptive system provides significantly more coverage than either the conventional or switched beam system [1], [2].

## 5. Relative Benefits / Tradeoffs of Switched Beam and Adaptive Array Systems

### 5.1. Integration

Switched beam systems are traditionally designed to retrofit widely deployed cellular systems. It has been commonly implemented as an add-on or appliqué technology that intelligently addresses the needs of mature networks.

### 5.2. Range / Coverage

Switched beam systems can increase base station range from 20 to 200 percent over conventional sectored cells, depending on environmental circumstances and the hardware/software used. The added coverage can save an operator substantial infrastructure costs and means lower prices for consumers. Also, the dynamic switching from beam to beam conserves capacity because the system does not send all signals in all directions. In comparison, adaptive array systems can cover a broader, more uniform area with the same power levels as a switched beam system.

### 5.3. Interference Suppression

Switched beam antennas suppress interference arriving from directions away from the active beam's center. Because beam patterns are fixed, however, actual interference rejection is often the gain of the selected communication beam pattern in the interferer's direction. Also, they are normally used only for reception because of the system's ambiguous perception of the location of the received signal (the consequences of transmitting in the wrong beam being obvious). Also, because their beams are predetermined, sensitivity can occasionally vary as the user moves through the sector.

Adaptive array technology currently offers more comprehensive interference rejection. Also, because it transmits an infinite, rather than finite, number of combinations, its narrower focus creates less interference to neighboring users than a switched-beam approach.

### 5.4. Spatial Division Multiple Access (SDMA)

Among the most sophisticated utilizations of smart antenna technology is SDMA, which employs advanced processing techniques to, in effect, locate and track fixed or mobile terminals, adaptively steering transmission signals toward users and away from interferers. This adaptive array technology achieves superior levels of interference suppression, making possible more efficient reuse of frequencies than the standard fixed hexagonal reuse patterns. *In essence, the scheme can adapt the frequency allocations to where the most users are located [19].*

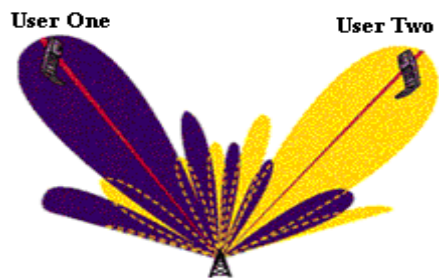

**Fig. 6 :  Fully Adaptive Spatial Processing, Supporting Two Users on the Same Conventional Channel Simultaneously in the Same Cell**

Because SDMA employs spatially selective transmission, an SDMA base station radiates much less total power than a conventional base station. One result is a reduction in network-wide RF pollution. Another is a reduction in power amplifier size. First, the power is divided among the elements, and then the power to each element is reduced because the energy is being delivered directionally. With a ten-element array, the amplifiers at each element need only transmit one-hundredth the power that would be transmitted from the corresponding single-antenna system [7].

Utilizing highly sophisticated algorithms and rapid processing hardware, spatial processing takes the reuse advantages that result from interference suppression to a new level. In essence, spatial processing dynamically creates a different sector for each user and conducts a frequency/channel allocation in an ongoing manner in real time.

Adaptive spatial processing integrates a higher level of measurement and analysis of the scattering aspects of the RF environment. Whereas traditional beam-forming and beam-steering techniques assume one correct direction of transmission toward a user, spatial processing maximizes the use of multiple antennas to combine signals in space in a method that transcends a one user-one beam methodology.

## 5.5. Adaptive Beam Forming and Algorithms

One of the foremost advantages offered by the software radio technology is flexibility. Because beam forming is implemented in software, it is possible to investigate a wide range of beam forming algorithms without the need to modify the system hardware for every algorithm. Consequently, researchers can focus their efforts on improving the performance of the beam forming algorithms rather than on designing new hardware, which can be a very expensive and time consuming process. A complete description of the RLS algorithm can be found in. This algorithm was chosen for its fast convergence rate and ability to process the input signal before demodulation. While the first reason is important especially when the environment is changing rapidly, the latter reason decreases the algorithm dependency on a specific air interface [5], [10] [19].

## 6. Spatial Structure Methods

The spatial structure is used to estimate the direction of arrivals (DOAs) of the signals impinging on the sensor array. The estimated directions of arrivals are then used to determine the weights in the pattern forming network. This is called beam forming. Spatial structure methods only exploit spatial structure and training signals and the temporal structure of the signals is ignored [20].

## 7. Adaptive Antennas

Adaptive antennas (AAs) are an array of antennas, which is able to change its antenna pattern dynamically to adjust to noise, interference and multipath. AAs are used to enhance received signals and may also be used to form beams for transmission [3]. AAs, unlike conventional antennas, confine the broadcast energy to a narrow beam. It optimizes the way that signals are distributed through space on a real time basis by focusing the signal to the desired user and "steering" it away from other users occupying the same channel in the same cell and adjacent or distant cell [4].

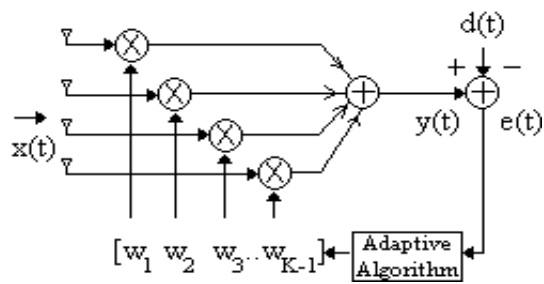

**Fig. 7 : Adaptive Beam Forming**

Na Yao demonstrated practically arraycomm smart antenna with only 4 elements and changed the shape and size of the radiation pattern sufficiently to prove the capability of smart antenna. He demonstrated a six-sector antenna pattern with four elements in each sector. The excitation of each element is varied to produce a shaped pattern for each sector, and those shapes vary sector by sector to produce the overall shaped radiation pattern for the cell [11].

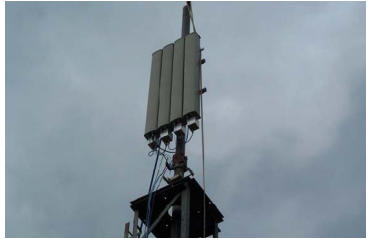

**Fig. 8 : Prototype Smart Antenna**

Arraycomm has been the forefront of developing smart antenna techniques and intellectual property for commercial cellular systems. *Intellicell technology is deployed in more than 90,000 commercial base station deployments worldwide but yet to be adopted in cellular communication network in INDIA.*

## 8. DISCUSSION

The above capability of adaptive / smart antennas clearly indicates without doubts that smart antenna can be easily replaced with traditionally used existing antennas (omni directional, sectored antenna with diversity concept). Use of adaptive antenna in existing systems will reduce power consumption and interference while enhancing spectral density in wireless system which is the dire need of wireless communication systems. Health hazard is being considered the main factor in RF communication which will also be taken care of by use of smart antenna as less RF pollution is created with the use of smart / adaptive antenna. Latest studies in INDIA indicates that most of the subsidies on diesel is being consumed by cellular vendors for running more than 4 lacs base stations in the country which is much more in comparison to farmers diesel consumption (this subsidy was meant for farmers and poor people only). The use of smart antenna will reduce diesel consumption in cellular communication drastically. The capabilities of smart antenna have been proved beyond doubt through analysis, simulation and deployment [12], [13], [14], [15], [16], [17], [18]. With the inclusion of nanotechnology devices the capability of adaptive antenna will increase manifold.

## 9. FUTURE SCOPE

Lot of work is in progress on smart antenna. Once nanotechnology antenna arrays are developed, it will be possible to incorporate smart antenna at handheld system too. As a result the performance of cellular systems will be enhanced manifold.

## 10. CONCLUSION

In conclusion to this paper "Smart Antenna" systems are the antennas with intelligence and the radiation pattern can be varied without being mechanically changed. With appropriate adaptive algorithms such as Recursive Least Square Algorithm (RLS) the beam forming can be obtained. As the system uses a DSP processor the signals can be processed digitally and the performance with a high data rate transmission and good reduction of mutual signal interference. The narrow beams get rid of interference, allowing many users to be connected with in the same cell at the same time using the same frequencies and can adapt the frequency allocation to where the most users are located. With adaptive beam forming, spectral efficiency of the cell could be multiplied at least ten times [6]. Smart antennas effectively reduce the power consumption which in turn avoids RF pollution, minimize health hazard and save scarce resource (diesel & foreign exchange). Indeed it has been argued that performance requirement of a future cellular communication   system cannot be made without the use of smart antennas.